\documentclass[aps,amsmath,twocolumn,amssymb,floatfixng,showpacs,superscriptaddress,footinbib]{revtex4-1}
\usepackage[T1]{fontenc}
\usepackage[utf8]{inputenc}

\pdfoutput=1
\usepackage[dvips]{graphics}
\usepackage{bm}
\usepackage{float}
\usepackage{epsfig}
\usepackage{enumerate}
\usepackage{subfigure}
\usepackage{color}
\usepackage{braket}
\usepackage{graphicx}
\usepackage{float}
\usepackage[colorlinks=true,linktoc=page,linkcolor=red,citecolor=blue,urlcolor=magenta]{hyperref}
\usepackage{amstext}
\usepackage{lipsum}
\usepackage{mathtools}
\usepackage{afterpage}
\usepackage{epstopdf} %converting to PDF
\usepackage{filecontents}
\usepackage{soul}

\newcommand{\myvec}[1]{\textbf{#1}}

\newcommand\bea{\begin{eqnarray}}
\newcommand\eea{\end{eqnarray}}
\newcommand\beq{\begin{equation}}  
\newcommand\eeq{\end{equation}}

\def\imag{i}

\usepackage[normalem]{ulem}
\usepackage{sidecap,tikz}
\DeclareRobustCommand{\orcidicon}{\hspace{-1.0mm}
	\begin{tikzpicture}
	\draw[lime, fill=lime] (0.0,0.0) 
	circle [radius=0.15] 
	node[white] {{\fontfamily{qag}\selectfont \tiny \,ID}};
	\draw[white, fill=white] (-0.0525,0.095) 
	circle [radius=0.007];
	\end{tikzpicture}
	\hspace{-3.0mm}
}
\foreach \x in {A, ..., Z}{\expandafter\xdef\csname orcid\x\endcsname{\noexpand\href{https://orcid.org/\csname orcidauthor\x\endcsname}
		{\noexpand\orcidicon}}
}

\begin{document}
\title{Topological characterization of special edge modes from the winding of relative phase}
\author{Sudarshan Saha}\email{sudarshan@iopb.res.in}\affiliation{Institute of Physics, Bhubaneswar- 751005, Odhisa, India}
\affiliation{Homi Bhabha National Institute, Mumbai - 400 094, Maharashtra, India}
\author{Tanay Nag\orcidB{}}
\email{tanay.nag@physics.uu.se}
\affiliation{Department of Physics and Astronomy, Uppsala University, Box 516, 75120 Uppsala, Sweden}
\author{Saptarshi Mandal\orcidC{}}\email{saptarshi@iopb.res.in}\affiliation{Institute of Physics, Bhubaneswar- 751005, Odhisa, India}
\affiliation{Homi Bhabha National Institute, Mumbai - 400 094, Maharashtra, India}

\begin{abstract}
%Various discrete symmetries play crucial roles in protecting the topological phases, characterized by appropriate topological invariants, where  bulk boundary correspondence is a novel phenomenon. Interestingly, in the absence of certain symmetries,  special edge modes can exist in many systems  indicating  the fact that  the symmetry-constrained topological invariant fails to explain the emergence  of these edge modes.

The symmetry-constrained topological invariant fails to explain the emergence  of the special edge modes when 
system does not preserve  discrete symmetries. 
The inversion or chiral symmetry broken SSH model 
is an example of one such system where
one-sided edge state with finite energy  appears at one end of the open  chain. 
To investigate whether this special edge mode is of topological origin or not, we introduce a concept of relative phase between the components of a two-component spinor and  define a winding number by the change of this relative phase  over the one-dimensional Brillouin zone. The relative phase winds non-trivially (trivially) in
accord with the presence (absence) of the one-sided edge mode inferring the bulk boundary correspondence. We extend this analysis to a two dimensional case where we characterize the   non-trivial phase, hosting  gapped one-sided edge mode, by the winding in relative phase only along a certain axis in the Brillouin zone. We demonstrate all the above findings from a generic parametric representation while topology is essentially determined by whether the underlying lower-dimensional projection includes or excludes the origin. Our study thus reveals a new paradigm of symmetry broken topological phases for future studies.

\end{abstract}

\date{\today}

%\pacs{insert PACS number here}
\maketitle

%\section{Introduction}
%%%%%%%%%%%%%%%%%%%%%%%%%%%%%%%%%%%%%%%%%%%%%%%%%%%%%%%%%%%%%%%%%%%%%%%%%%%%%%%%%%%%%%%%%%%%%%%%%%
%\section{Introduction}
%%%%%%%%%%%%%%%%%%%%%%%%%%%%%%%%%%%%%%%%%%%%%%%%%%%%%%%%%%%%%%%%%%%%%%%%%%%%%%%%%%%%%%%%%%%%%%%%%%
\textcolor{red}{Introduction}---
Over the last two decades, unprecedented progress has been made in understanding and exploring  various novel topological states \cite{hasan-2010,qi-2011}. It has been realized that a given topological system must be described by a set of 
underlying criteria such as  symmetry \cite{altland-1997,ryu-2010,chiu-2016,Hasan-2010},  bulk topological invariants and 
characteristic edge states establishing the bulk-boundary correspondences \cite{okugawa-2022,saha-2021}. The topological characterisation of many
condensed matter system based on the  geometric phase \cite{berry-1984,Jak-1989,xiao-2010,benalcazar-2017} established a paradigm of itself.
The geometric phase or Berry phase, derived from Berry curvature, is intimately connected to the
topological invariant  namely, Chern number and historically the Chern insulator is the first generation of topological phases for time reversal symmetry broken systems \cite{hatsugai-1993,haldane-1988,kane-mele-2005,kane-mele-2005a,prodan-2009,wang-2015,yu-2010}.

Even with the increased focus on symmetry-constrained topological characterisations \cite{altland-1997,ryu-2010,chiu-2016}, there are some exceptions which are not explained by the previously established topological characterisations.
Interestingly, it has been noticed that various symmetry broken systems such as inversion symmetry (IS) broken Su-Schrieffer-Heeger (SSH) model  in one dimension (1D) \cite{Kawarabayashi-2021,shen-book}, $C_3$-symmetry broken
Haldane model in two dimension (2D) \cite{wang-2021,saha-2023}, host unconventional edge states but the existing topological invariants are  incapable to characterize these  phases.  It would not be prudent to mark these phases as non-topological where the such special edge modes come from trivial origin \cite{Ryu-2002}.  This leads to an open question regarding the topological origin of such edge modes  that we seek to answer in this work. To be precise, does there exist an alternative way to define the topological invariants for IS broken  systems (or any symmetry broken system in general) that allows for a proper justification behind the emergence of such edge modes and establishes the topological character of these systems? To the best of our knowledge, this is the first attempt to  understand the unconventional/special edge modes from the bulk property of the systems. \\

In this direction, we start with a generic parametric approach to set up the models. The first model reduces to the  widely investigated one-dimensional  SSH model \cite{ssh-1979,ssh-1980,rice-mele-1982} where the 
 IS is broken  by the presence of  mass terms with opposite signs for each sub-lattices. This results in the finite energy one-sided edge states. 
In momentum  `$k$' space  the wave-function  is described by a two-component spinor. To  illustrate our idea,  we first define an appropriate relative phase between the spinor components and  consider the  winding
of this relative phase over the Brillouin zone (BZ) in `$k$'.  We find that non-trivial (trivial) winding can successfully explain the presence (absence) of the one-sided edge modes. 
%To the best of our  knowledge,  this is the first of its kind a demonstration showing the edge modes of IS broken SSH model might have a  topological origin.
To strengthen our finding further in the higher dimension, we construct a 2D IS broken model, with anisotropic hopping, out of the parametric realization and find  that the above characterisation holds true. Importantly, the topological (trivial) winding essentially captures the inclusion (exclusion) of origin for the lower-dimensional projection of the higher-dimensional parametric representation irrespective of the choice of the models.

%%%%%%%%%%%%%%%%%%%%%%%%%%%%%%%%%%%%%%%%%%%%%%%%%%%%%%%%%%%%%%%%%%%%%%%%%%%%%%%%%%%%%%%%%%%%%%%%%%
%\section{Basic of topology of relative phase}
%\label{basic}
%%%%%%%%%%%%%%%%%%%%%%%%%%%%%%%%%%%%%%%%%%%%%%%%%%%%%%%%%%%%%%%%%%%%%%%%%%%%%%%%%%%%%%%%%%%%%%%%%%
\textcolor{red}{Parametric representation and relative phase winding}---
Here we elaborate  the  concept of possible topological characterization using the notion of relative phase for a two-component wave-function and its winding over BZ. We appropriately define the winding  as the topological marker. To describe nontrivial topology with tight-binding Hamiltonian we must have at least a two-component wave-function. We start with  two states per lattice site and corresponding basis states  are represented by $\Ket{a_1}, \Ket{a_2}$. Any  normalized state can be expressed as $\Ket{\Psi}_k = \psi_{1,k}  \Ket{a_1} + \psi_{2,k} \Ket{a_2}$
where $k$ is a continuous parameter with a certain periodicity.
In general $\psi_{1,k}$ and $\psi_{2,k}$ are complex numbers and their relative phase $\phi_R = \arg(\psi_1) - \arg(\psi_2)$, is a gauge independent quantity; here we omit the obvious $k$ subscript.
Considering $\psi_{1} = |\psi_{1}|  e^{\imag\theta_{1}}, \psi_2 =|\psi_{2}| e^{\imag\theta_{2}}$ and one can identify the measure of winding $\omega_{12}$, associated with  the wave-function, as follows $\omega_{12}=  {\rm Im}[\int_{0}^{2\pi} dk \: \psi_1^*\psi_2 \partial_k (\psi_1\psi_2^*)]/2 \pi
= \Delta \phi_R/2\pi$
where $\phi_R= (\theta_1 - \theta_2)$. Importantly, $\omega_{12}$  defines the winding number which is quantized to zero or finite integers for trivial and
non-trivial topological phases, respectively. 

%%%%%%%%%%%%%%%%%%%%%%%%%%%%%%%%%%%%%%%%%%%%%%%%%%%%%%%%%%%%%%%%%%%%%%%%%%%%%%%%%%%%%%%%%%%%%%%%%%
\begin{figure}[!ht]
    \begin{center}
        \includegraphics[scale=0.24]{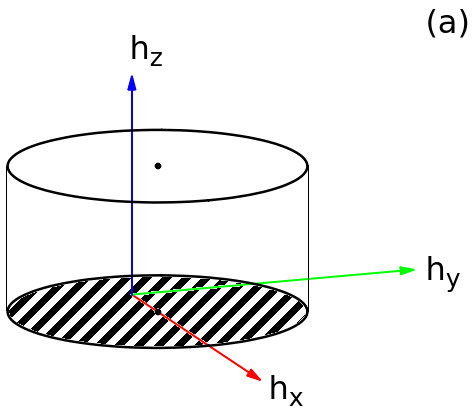}
        \hspace{23pt}
        \includegraphics[scale=0.24]{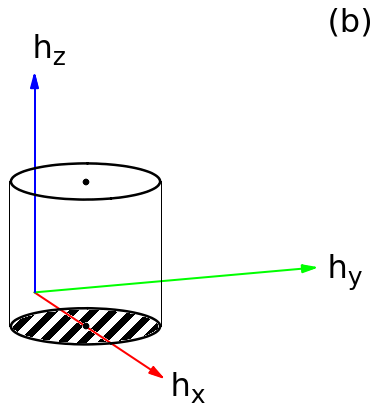}
        \caption{Parametric representation for 1D: A circle  represented in the $h_x$-$h_y$-$h_z$ plane. 
        The upper empty circles depict the contour formed by $\myvec{d}$ as we vary $k$.  The striped-filled circles  below represent the projection of the empty circles on the $h_x$-$h_y$ plane.  (a) and (b)  correspond to non-trivial and trivial phases, where striped-filled circle includes and excludes the origin with parameters $(v, r, M) = (0.5, 1.0, 1.0)$ and $(1.0, 0.5, 1.0)$, respectively.}
        \label{Fig:visualizeBulkHamiltonian}
    \end{center}
\end{figure}

%%%%%%%%%%%%%%%%%%%%%%%%%%%%%%%%%%%%%%%%%%%%%%%%%%%%%%%%%%%%%%%%%%%%%%%%%%%%%%%%%%%%%%%%%%%%%%%%%%

Given a generalized two-level Hamiltonian of the form $H=\sum_{\alpha}h_{\alpha} \sigma_{\alpha}$ with $\sigma_{\alpha}$ representing the Pauli matrices and $\alpha=x,y,z$. The valence band wave-function can be written as $\Ket{\psi_-} = \frac{1}{N_-} \left(h_z-h,h_x+ i h_y \right)^T$  where $N_- = [2h(h - h_z)]^{1/2}$, $h=(h^2_x+h^2_y+h^2_z)^{1/2}$, $T$ denotes the transpose operation and for simplicity we suppress the obvious $k$ dependence in the relevant quantities. Given the parametric dependence of $h_{\alpha}$ on $k$,  we find that the projection of the tip of the vector $\myvec{d}= \sum _{\alpha}h_{\alpha} \hat{e}_{\alpha}$, with $\hat{e}_{\alpha}$ representing unit vector, 
on the $h_x$-$h_y$ plane may or may not encircle the origin  as shown in Figs. \ref{Fig:visualizeBulkHamiltonian}, \ref{Fig:visualizeBulkHamiltonianIn2d}. One can thus find $\phi_R = \arg(h_x + i h_y)$. It is now evident that the change in relative phase is directly connected with the conventional notion of inclusion or exclusion of origin in the $h_x$-$h_y$ plane. Note that $\omega_{12}$ reduces to the winding number when $h_z=0$ for the system preserving the chiral symmetry \cite{Goerbig-2014}. 

%%%%%%%%%%%%%%%%%%%%%%%%%%%%%%%%%%%%%%%%%%%%%%%%%%%%%%%%%%%%%%%%%%%%%%%%%%%%%%%%%%%%%%%%%%%%%%%%%%
\begin{figure}
    \centering
    \includegraphics[scale=0.3]{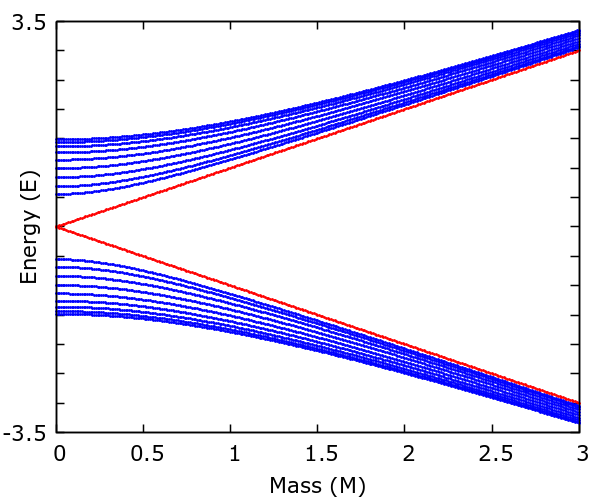}
    \caption{We show the energy spectrum of 1D model as a function of sub-lattice mass term
    with  open boundary conditions.   The finite energy one-sided  edge modes, appearing for finite values of $M$, are shown in red. We choose $(v, r) = (0.5, 1.0)$ and $N=10$ lattice sites. }
    \label{fig3}
\end{figure}
%%%%%%%%%%%%%%%%%%%%%%%%%%%%%%%%%%%%%%%%%%%%%%%%%%%%%%%%%%%%%%%%%%%%%%%%%%%%%%%%%%%%%%%%%%%%%%%%%%

%%%%%%%%%%%%%%%%%%%%%%%%%%%%%%%%%%%%%%%%%%%%%%%%%%%%%%%%%%%%%%%%%%%%%%%%%%%%%%%%%%%%%%%%%%%%%%%%%%
%\section{Example with inversion broken 1d SSH model}
%\label{1dmodel}
%%%%%%%%%%%%%%%%%%%%%%%%%%%%%%%%%%%%%%%%%%%%%%%%%%%%%%%%%%%%%%%%%%%%%%%%%%%%%%%%%%%%%%%%%%%%%%%%%%

%%%%%%%%%%%%%%%%%%%%%%%%%%%%%%%%%%%%%%%%%%%%%%%%%%%%%%%%%%%%%%%%%%%%%%%%%%%%%%%%%%%%%%%%%%%%%%%%%%
\begin{figure}[!ht]
    \begin{center}
        \includegraphics[scale=0.32]{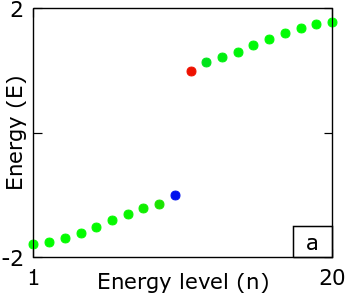}
        \hspace{-5pt}
        \includegraphics[scale=0.32]{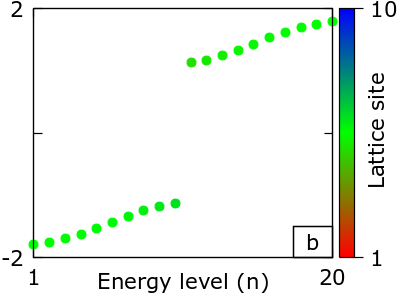}
        \caption{We demonstrate the energy spectrum of 1D model for  $(v, r, M) = (0.5, 1.0, 1.0)$ and $(1.0,0.5,1.0)$ in (a) and (b), respectively, with open boundary condition where the color bar represents the  localization profile.  In the non-trivial phase, the postive (negative) energy mid-gap state is localized at the left edge $N=1$ (right edge $N=10$).  }
        \label{Fig:bandStructureFiniteGeometry}
    \end{center}
\end{figure}
%%%%%%%%%%%%%%%%%%%%%%%%%%%%%%%%%%%%%%%%%%%%%%%%%%%%%%%%%%%%%%%%%%%%%%%%%%%%%%%%%%%%%%%%%%%%%%%%%%

%%%%%%%%%%%%%%%%%%%%%%%%%%%%%%%%%%%%%%%%%%%%%%%%%%%%%%%%%%%%%%%%%%%%%%%%%%%%%%%%%%%%%%%%%%%%%%%%%%
\begin{figure}[!ht]
    \centering
    \includegraphics[scale=0.32]{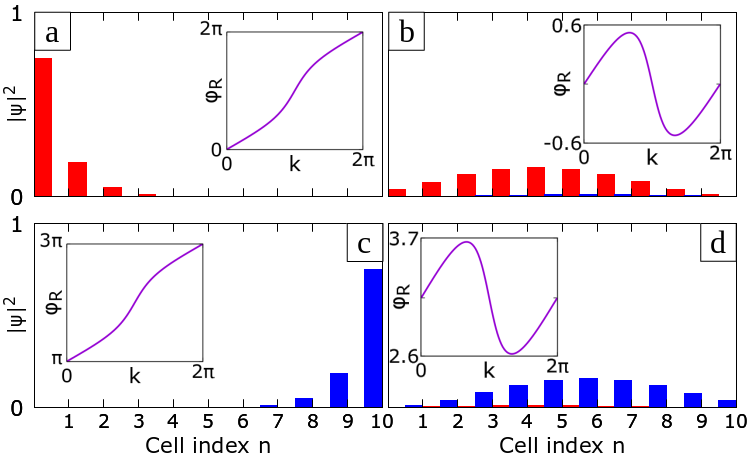}
    \caption{We show the probability density of the positive and negative energy mid-gap states i.e., lowest conduction and highest valence bands in (a) and (c), respectively, for the topological phase of 1D model under open boundary condition. The insets in (a) and (c) show the variation of relative phase  $\phi_R$, computed from the bulk Hamiltonian,  over $k$ for the valence and conduction band, respectively.  We repeat (a,c) in (b,d) for the trivial phase where neither one-sided edge mode nor non-trivial winding in $\phi_R$ is observed. We consider the same parameter set as described in Fig. 
\ref{Fig:bandStructureFiniteGeometry}.
}
\label{Fig:relativePhase}
\end{figure}
%%%%%%%%%%%%%%%%%%%%%%%%%%%%%%%%%%%%%%%%%%%%%%%%%%%%%%%%%%%%%%%%%%%%%%%%%%%%%%%%%%%%%%%%%%%%%%%%%%

\textcolor{red}{Application to 1D model}---
Here we implement our formalism in 1D to test the definition of winding $\omega_{12}$. We consider $ h_x= v + r \cos k, h_y= r \sin k$, and $h_z=M$ such that 
parametric representation yields a circle $(h_x-v)^2 + h_y^2 = r^2$ whose centre is located at $(h_x,h_y,h_z)=(v,0,M)$ (see Fig. \ref{Fig:visualizeBulkHamiltonian}). This corresponds to the IS broken SSH model with dimerized hopping amplitudes being weak and strong in successive bonds and finite sub-lattice mass terms with  opposite signs. The model preserves time reversal symmetry (TRS),   generated by $K$,  where $K$ denotes the complex conjugation. Importantly, the chiral symmetry and particle hole symmetry (PHS), generated by $\sigma_z$ and $\sigma_z K$, are respectively broken for $M\ne 0$ resulting in the breakdown of chiral symmetry-constrained winding number \cite{shen-book,Cayssol-2021,Fuchs-2021,Brzezicki-2020,Shiozaki-2015}.

For $M=0$, IS generated by $\sigma_x$ is preserved; there exist two mid-gap states at zero energy for $|v|<|r|$; two different sub-lattices, existing at two opposite  
ends of the open chain get simultaneously populated by each of such states. The degeneracy of these states is lifted once IS is broken with $M\ne 0$ leading to the fact that positive and negative energy states are separately  localized at two different sub-lattices (see Figs.   \ref{fig3} and \ref{Fig:bandStructureFiniteGeometry}).  These  finite energy special edge modes are referred to as the one-sided edge modes.
These states gradually approach the bulk states as $M$ increases while their localization properties remain unaltered.  Note that only one end of the chain is occupied for half-filled case unlike the IS preserved case where both the ends are occupied.     For $|v|>|r|$, there is no such one-sided mid-gap state in the open chain.

Following the description of geometric phase,
the winding number $\nu=\frac{1}{2\pi i} \int^{2\pi}_0 dk \frac{d}{dk} {\rm{log}} z(k)$ where $z(k)=h_x(k)-i h_y(k)$, computed for $M=0$, is found to be $1$ and $0$ for $|v|<|r|$  and $|v|>|r|$, respectively. The emergence of zero-energy edge modes for $|v|<|r|$ is attributed to the bulk-boundary correspondence.  For finite $M$, the above topological characterization becomes invalidated even though there exist finite energy one-sided edge modes. This indicates an apparent breaking of bulk-boundary correspondence. To this end,  we show that the winding number $\omega_{12}$, defined by the difference in relative phase over 1D BZ,  serves as a useful bulk topological order parameter for finite $M$. Interestingly,  $\phi_R$ takes two different (identical) values of phase, separated by $2\pi$ ($0$), at the two ends of the  BZ for  $|v|<|r|$ ($|v|>|r|$). This indicates that $\phi_R$ winds over the BZ non-trivially (trivially) for $|v|<|r|$ ($|v|>|r|$) leading to $\omega_{12}=1$ ($\omega_{12}=0$). This observation goes hand in hand with the presence and absence of the one-sided edge modes as shown in  Fig. \ref{Fig:relativePhase}. 
 Importantly, the tip of the vector $\vec{d}$ covers a circle whose projection on $h_x$-$h_y$ plane includes (excludes) origin for  $|v|<|r|$ ($|v|>|r|$)
as shown in Fig. \ref{Fig:visualizeBulkHamiltonian} even with finite values of $M$. 
Thus we are able to show conclusively that for finite $M$, the relative phase winding can signal the emergence of special edge states.

%%%%%%%%%%%%%%%%%%%%%%%%%%%%%%%%%%%%%%%%%%%%%%%%%%%%%%%%%%%%%%%%%%%%%%%%%%%%%%%%%%%%%%%%%%%%%%%%%%
%\section{Application to 2D}
%\label{2dmodel}
%%%%%%%%%%%%%%%%%%%%%%%%%%%%%%%%%%%%%%%%%%%%%%%%%%%%%%%%%%%%%%%%%%%%%%%%%%%%%%%%%%%%%%%%%%%%%%%%%%

\textcolor{red}{Application to 2D model}---
After the successful demonstration of the topological invariant and the associated bulk-boundary correspondence in 1D, we now explore its application to a 2D problem.  We now have two cyclic parameters namely, $k_x$ and $k_y$ while for 1D, we have only one cyclic parameter. 
Without loss of generality, we adopt the parametric representation as $h_x = v + r_1 \cos(k_x) + r_2 \cos(k_x) \cos(k_y), h_y = r_1 \sin(k_x) + r_2 \sin(k_x) \cos(k_y)$,  and $ h_z= M + r_2 \sin(k_y)$ yielding a torus $\left(r_1-\sqrt{(h_x-v)^2+h_y^2}\right)^2+h_z^2=r_2^2$ whose center is located at $(h_x,h_y,h_z)=(v,0,M)$ (see Fig. \ref{Fig:visualizeBulkHamiltonianIn2d}). Here, $r_1$ and $r_2$ represent the larger and smaller radius.

%%%%%%%%%%%%%%%%%%%%%%%%%%%%%%%%%%%%%%%%%%%%%%%%%%%%%%%%%%%%%%%%%%%%%%%%%%%%%%%%%%%%%%%%%%%%%%%%%%
\begin{figure}[!ht]
    \begin{center}
        \includegraphics[scale=0.253]{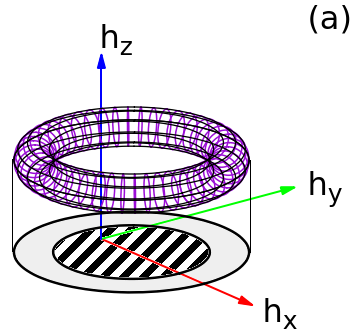}
		\hspace{45pt}
        \includegraphics[scale=0.25]{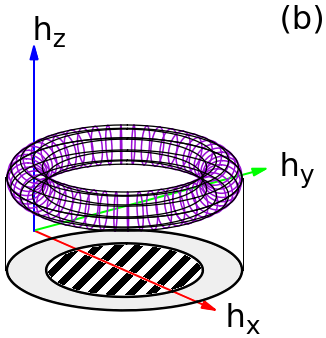}
        \caption{Parametric representation for 2D: A torus is represented on the $h_x$-$h_y$-$h_z$ plane. 
        The torus is formed by the $\myvec{d}$ as we vary $k_x$ and $k_y$ where the projection on $h_x$-$h_y$ plane is captured by two concentric circles of radius $r_1$ and $r_2$ with $r_1>r_2$. The striped-filled (grey-shaded) part designates the inside (annular) region, associated with the projection.  In (a), the projection  contains  origin inside the inner circle of radius $r_2$ as $|v|<|r_1-r_2|$. In (b), the projection  does not contain origin which is outside the outer circle as $|v|>|r_1+r_2|$. We consider $(v,r_1,r_2,M) = (0.5,1.0,0.2,1.0)$ and  $(1.5,1.0,0.2,1.0)$, respectively, for (a) and (b).}
        \label{Fig:visualizeBulkHamiltonianIn2d}
    \end{center}
\end{figure}
%%%%%%%%%%%%%%%%%%%%%%%%%%%%%%%%%%%%%%%%%%%%%%%%%%%%%%%%%%%%%%%%%%%%%%%%%%%%%%%%%%%%%%%%%%%%%%%%%%

This representation can be regarded as a tight-binding Hamiltonian on a square lattice with two sub-lattices in a given unit cell. The intracell hopping amplitude is $v$ while $r_{1,2}$ denotes the intercell hopping amplitude; $M$ refers to the sub-lattice mass term that causes the  breaking IS generated by  $\sigma_x$.
One of the hopping is found to be imaginary resembling with the  flux-modulated hopping of the Haldane model; this results in the breaking of TRS generated by $K$.   
This model breaks chiral symmetry and PHS, generated by $\sigma_z$ and $\sigma_z K$, respectively. The projection of the torus on the $h_x$-$h_y$ plane includes (excludes) origin for $|v|<|r_1-r_2|$ ($|v|>|r_1+r_2|$), irrespective of the values of $M$.  On the other hand, $r_1 \ne r_2$ refers to an anisotropic case that is characteristically different from the isotropic case $r_1=r_2$ where the torus reduces to a sphere in the parametric space. One can check that the zero- (finite-) energy conventional edge modes exist for the isotropic case with $M=0$ ($M\ne 0$) as a signature of the topological phase. This is similar to the Chern insulator phases.

\begin{figure}[!ht]
    \begin{center}
        \includegraphics[scale=0.265]{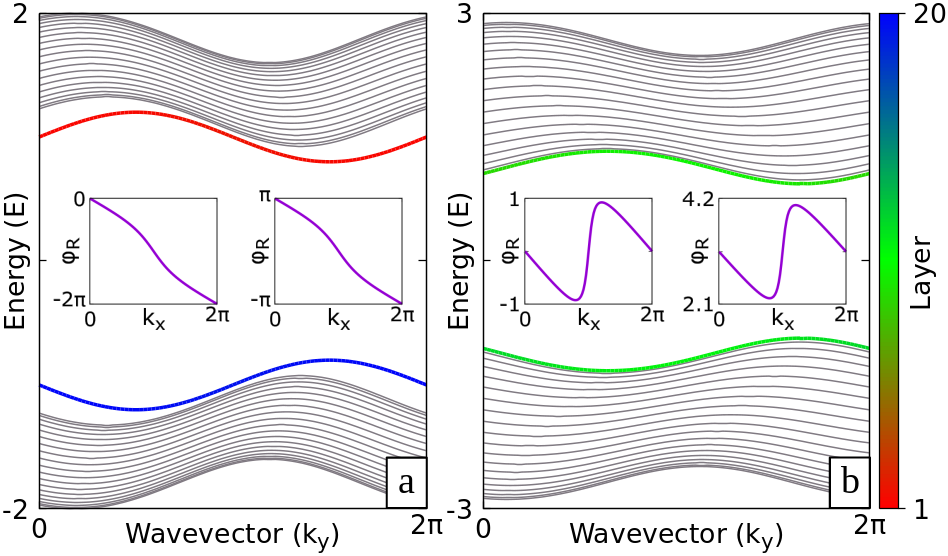}
        \includegraphics[scale=0.265]{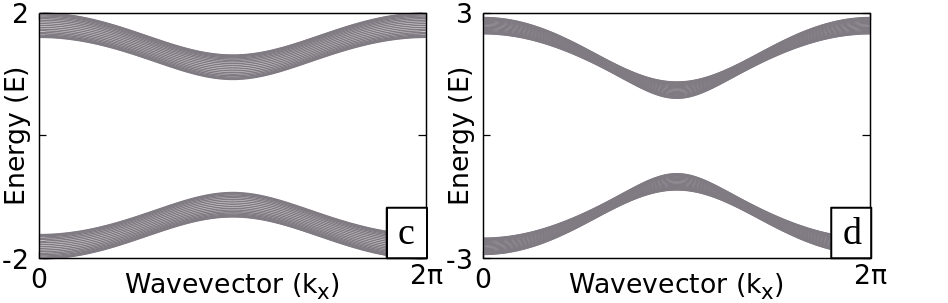}
        \caption{We show ribbon geometry band structure of 2D model, considering 
        open boundary condition along $x$ and $y$-direction in (a) and (c), respectively, that indicates the   one-sided unconventional edge modes at finite energy as a signature of the topological phase in the 2D model. Inset shows the non-trivial winding of the relative phase, computed from the bulk Hamiltonian with $k_y=0$, as a function of $k_x$  for both the valence and conduction bands. We repeat (a,c) in (b,d) for the trivial phase where  one-sided edge states and non-trivial winding do not show up.  
        We consider the same parameter set as described in Fig. \ref{Fig:visualizeBulkHamiltonianIn2d}.}
        \label{ribbon-non-trivial}
    \end{center}
\end{figure}

We here examine the anisotropic limit, representing the torus in the parameter space 
We restrict ourselves to the parameter window $M>r_2$ that ensures $h_z$ to be always finite. We cut the strong (weak) bond i.e., open boundary condition along $x$($y$)-direction to compute ribbon geometry band structure with $k_y$ ($k_x$) where we find finite energy mid-gap states are localized only over the boundaries along $x$-direction at $x=\pm L$ (see Figs. \ref{ribbon-non-trivial}(a,c)). Interestingly, we do not find any states localized over the boundaries along $y$-direction at $y=\pm L$ suggesting the fact that these 1D edge modes exhibit special localization properties unlike the conventional  1D edge modes in 2D Chern insulator. Similar to the earlier 1D case, we find the one-sided unconventional edge states with  finite energy for the 2D case as well. 
On the other hand, by exploiting the bulk Hamiltonian, we find how the relative phases of the conduction and valence band non-trivially wind
along $k_x$ for values of $k_y$. This results in the winding number $\omega_{12}$ to take finite value for $|v|<|r_1-r_2|$. The non-trivial winding of the relative phase is always (never) observed with $k_x$ ($k_y$) irrespective of the values of $k_y$ ($k_x$) encoding the underlying bulk information for the one-sided edge modes.  In the trivial phase, for $|v|>|r_1+r_2|$, we do not find any one-sided edge modes under open boundary condition and consequently, there exists no non-trivial winding for the bulk bands (see Figs. \ref{ribbon-non-trivial}(b,d)). This further establishes the bulk-boundary correspondence for the 2D case similar to that of the 1D case. We can comment that the above findings are expected to hold for $M<r_2$, however, the detailed analysis we leave for future study.

One can find another parameter window $|r_1-r_2|<|v|<|r_1+r_2|$ where the projection, associated with the annular region of the  torus,  includes the origin rather than inner circle of the torus enclosing the origin. There is a window of $k_y$, symmetrically placed around $k_y=0$ and bounded  by $ k_y< |{\rm arccos}(|v-r_1|/|r_2|)|$, for which the relative phase winds  non-trivially across $k_x$.  Once $|v|$ reaches $|r_1-r_2|$ ($|r_1+r_2|$), the  relative phase winding  is  present (absent) as a function of $k_x$ for all values of $k_y$. This refers to a unique crossover region instead of a phase boundary between the trivial and non-trivial phase. In this region, the mid-gap states are not localized at $x=\pm L$ irrespective of the choice of $k_y$ when the ribbon geometry band dispersion is studied.

%%%%%%%%%%%%%%%%%%%%%%%%%%%%%%%%%%%%%%%%%%%%%%%%%%%%%%%%%%%%%%%%%%%%%%%%%%%%%%%%%%%%%%%%%%%%%%%%%%
%%%%%%%%%%%%%%%%%%%%%%%%%%%%%%%%%%%%%%%%%%%%%%%%%%%%%%%%%%%%%%%%%%%%%%%%%%%%%%%%%%%%%%%%%%%%%%%%%%

%%%%%%%%%%%%%%%%%%%%%%%%%%%%%%%%%%%%%%%%%%%%%%%%%%%%%%%%%%%%%%%%%%%%%%%%%%%%%%%%%%%%%%%%%%%%%%%%%%	
%\section{Discussion}
%\label{discussion}
%%%%%%%%%%%%%%%%%%%%%%%%%%%%%%%%%%%%%%%%%%%%%%%%%%%%%%%%%%%%%%%%%%%%%%%%%%%%%%%%%%%%%%%%%%%%%%%%%%
\textcolor{red}{Conclusion and discussion}---
To summarise,  we introduce a new definition of topological invariant based on the winding of the relative phase for a two-level systems.
We are motivated by the fact that there exist symmetry-broken intriguing topological phases, hosting special edge modes exist, while the symmetry-constrained conventional topological invariant is unable to characterize them. We adopt a generic parametric representation from which 1D and 2D lattice models can be inferred to verify the prescribed methodology.    
We find unusual topological phases, hosting gapped unconventional  one-sided edge modes under open boundary condition, for the  IS broken SSH model and anisotropic hopping model in 1D and 2D, respectively. The lower-dimensional projection of the parent parametric representation is able to identify these topological (trivial) phases by examining the inclusion (exclusion) of the origin. The relative phase, computed from the wave-function corresponding to the bulk Hamiltonian, exhibits non-trivial  winding when there exist one-sided edge states.  We further note that the present method can  characterize the topological phases for IS broken as well as preserved models  in a unified manner as long as the wave-function remains non-singular.

 This results in opening up new  possibilities to search for  topological classifications and bulk-boundary correspondence beyond present symmetry criteria. Note that, trivial and topological phases are not separated by a gapless critical line due to the presence of the finite mass term. Therefore, the notion of the topological phase boundary is also not evident except for the fact that the nature of the  relative phase winding changes. In terms of the lower-dimensional projection, we represent the contour of the  IS preserved instantaneous Hamiltonian. The inclusion and exclusion of the origin for this projected Hamiltonian  are thus associated with a bulk gap closing  when the above contour crosses the origin.

One obvious future problem is when there are more than two components in the wave-function  we need to deal with various combinations of them which will lead to more than one relative phase. We think the way the spin-Chern number has been introduced to characterize the topology of a spin-full system, one may use the spin-spectrum gap to identify two valance band states and then individually identify the winding of various
combinations of relative phases. However, the scope of such a study remains open for the future. The most significant issue  that requires a great deal of attention is the order of these topological phases as they are not classified on the basis of relative phase.

%%%%%%%%%%%%%%%%%%%%%%%%%%%%%%%%%%%%%%%%%%%%%%%%%%%%%%%%%%%%%%%%%%%%%%%%%%%%%%%%%%%%%%%%%%%%%%%%%%
%\section{Acknowledgement}

\textcolor{red}{Acknowledgement}---
SS acknowledges SAMKHYA (High-Performance Computing Facility provided by the Institute of Physics, Bhubaneswar) for the numerical computation.
SM acknowledges support from the ICTP through the Associate's Programme (2020-2025).
%%%%%%%%%%%%%%%%%%%%%%%%%%%%%%%%%%%%%%%%%%%%%%%%%%%%%%%%%%%%%%%%%%%%%%%%%%%%%%%%%%%%%%%%%%%%%%%%%%

%%%%%%%%%%%%%%%%%%%%%%%%%%%%%%%%%%%%%%%%%%%%%%%%%%%%%%%%%%%%%%%%%%%%%%%%%%%%%%%%%%%%%%%%%%%%%%%%%%
\bibliography{bibfile}{}

%\appendix
%%%%%%%%%%%%%%%%%%%%%%%%%%%%%%%%%%%%%%%%%%%%%%%%%%%%%%%%%%%%%%%%%%%%%%%%%%%%%%%%%%%%%%%%%%%%%%%%%%

\end{document}